\documentclass[twocolumn,aps,prd,nofootinbib]{revtex4-2}
 \usepackage{graphicx,amssymb}
 \usepackage[10pt]{extsizes}
\usepackage[utf8]{inputenc}
\usepackage[T2A]{fontenc}
\usepackage[strict]{changepage}
\usepackage[main=english]{babel}
\usepackage{amsmath} 
\usepackage{amssymb}
\usepackage{graphicx}
\usepackage{subfigure}
\usepackage{bm}
\usepackage{indentfirst}
\usepackage{placeins}
\usepackage[left=20mm, top=20mm, right=20mm, bottom=20mm, nohead, footskip=7mm]{geometry}
 \usepackage[utf8]{inputenc}
 
\usepackage[urlcolor=blue,colorlinks,breaklinks=true]{hyperref}

\PassOptionsToPackage{obeyspaces,spaces,hyphens}{url}

 \usepackage{amsmath}
 
\begin{document}
 	%My commands
 	\def\half{{1\over2}}
 	\def\shalf{\textstyle{{1\over2}}}
 	
 	\newcommand\lsim{\mathrel{\rlap{\lower4pt\hbox{\hskip1pt$\sim$}}
 			\raise1pt\hbox{$<$}}}
 	\newcommand\gsim{\mathrel{\rlap{\lower4pt\hbox{\hskip1pt$\sim$}}
 			\raise1pt\hbox{$>$}}}

\newcommand{\be}{\begin{equation}}
\newcommand{\ee}{\end{equation}}
\newcommand{\bq}{\begin{eqnarray}}
\newcommand{\eq}{\end{eqnarray}}
\newcommand{\vv}{\bar{v}}
\newcommand{\cc}{{\tilde c}}

\title{Ultra-high frequency gravitational waves from cosmic strings with friction}

\author{S. Mukovnikov}
\email[Electronic address: ]{sergei.mukovnikov@astro.up.pt}
\affiliation{Centro de Astrof\'{\i}sica da Universidade do Porto, Rua das
Estrelas, 4150-762 Porto, Portugal}
\affiliation{Instituto de Astrof\'{\i}sica e Ci\^encias do Espa\c co,
CAUP, Rua das Estrelas, 4150-762 Porto, Portugal}
\affiliation{Departamento de F\'{\i}sica e Astronomia, Faculdade de Ci\^encias, Universidade do Porto, Rua do Campo Alegre 687, PT4169-007 Porto, Portugal}

\author{L. Sousa}
\email[Electronic address: ]{lara.sousa@astro.up.pt}
\affiliation{Centro de Astrof\'{\i}sica da Universidade do Porto, Rua das
Estrelas, 4150-762 Porto, Portugal}
\affiliation{Instituto de Astrof\'{\i}sica e Ci\^encias do Espa\c co,
CAUP, Rua das Estrelas, 4150-762 Porto, Portugal}
\affiliation{Departamento de F\'{\i}sica e Astronomia, Faculdade de Ci\^encias, Universidade do Porto, Rua do Campo Alegre 687, PT4169-007 Porto, Portugal}

\date{\today}

\date{\today}
\begin{abstract}
We include the effect of the frictional force caused by interactions between cosmic strings and the particles of the background plasma in the computation of the stochastic gravitational wave background generated by cosmic string loops. Although our results show that friction leads to a partial suppression of the emission of gravitational radiation by cosmic string loops, we also find that loop production is very intense in the early stages of the Kibble regime. We show that, in many instances, this leads to a prominent signature of friction in the ultra-high frequency range of the spectrum, in the form of a secondary peak. The signature of friction is not only sensitive to cosmic string properties, but also to the initial conditions of the network and its surroundings. A detection of this signature would then allow us to extract information about the physics of the early universe that cannot be uncovered when probing the rest of the stochastic gravitational wave background spectrum.
\end{abstract}

\maketitle

\section{Introduction}

We are swiftly moving into the era of Gravitational Wave (GW) Astronomy. The LIGO-Virgo Collaboration has observed, over the past years, numerous compact object coalescence events~\cite{LIGOScientific:2016sjg, LIGOScientific:2017vwq} and very recently major millisecond pulsar timing arrays have announced the detection of a Stochastic Gravitational Wave Background (SGWB)~\cite{NANOGrav:2023gor, EPTA:2023fyk, Reardon:2023gzh, Xu:2023wog} in the nanohertz frequency range. Gravitational radiation, unlike electromagnetic radiation, travels freely through spacetime even in the very early stages of the evolution of the universe and may then allow us to study previously undetected sources. Cosmic strings may be one such source. The early universe is expected to have undergone a series of symmetry-breaking phase transitions that may lead to the production of these linelike topological defects. Cosmic strings are generally expected to survive until the present time and, since they concentrate a significant amount of energy and move with relativistic velocities, they may potentially leave different observational signatures~\cite{M_B_Hindmarsh_1995, Vilenkin:2000jqa}. These remnants of the very early universe may then allow us to probe particle physics up to very high energies, even beyond the reach of current and future collider experiments.

Although standard observational probes have failed to detect cosmic strings, with the onset of GW Astronomy, we have a new promising way to probe them: the SGWB they generate~\cite{hogan,accetta}. The SGWB generated by cosmic string networks is mainly sourced by the emissions of the closed loops of string that are continuously produced in string interactions. In the initial stages of the evolution of the network, when the universe is quite dense, strings are damped by the frictional force caused by the frequent interactions with particles of the surrounding plasma~\cite{PhysRevD.43.1060,Garriga,Martins:1996jp}. Studies of the cosmic string SGWB~\cite{hogan,accetta,Blanco-Pillado:2013qja,Sanidas:2012ee,Sousa:2013aaa,auclair}  often assume that loops created in this friction-dominated era do not provide a significant contribution to this spectrum and that GW production only starts once the universe has rarified enough for friction to become irrelevant for string dynamics. However, although friction should in fact lead to a decrease of the GW emission of loops, string networks are expected to be quite dense in the friction era, which means that many loops should be produced. As a result, the contribution of the loops created in the friction era may not, in fact, be negligible as is usually assumed. Here, we study the impact of friction on the GW emission of loops and on the number of loops produced, with the objective of computing the contribution of the loops created in the friction era to the SGWB. We show that friction may lead to a distinct signature in the high frequency range of the spectrum, in the form of a secondary peak, and show that its properties are dependent on cosmic string properties and on the properties of the background plasma. The SGWB generated by cosmic strings may then extend further into the ultra-high frequency range of the GW spectrum than previously anticipated, allowing us then to look further into the early stages of the evolution of the universe.

This paper is organized as follows. In Sec.~\ref{sec:network-evolution}, we introduce the Nambu-Goto equations of motion for cosmic strings with friction and outline how these may be used to derive a model to describe the cosmological evolution of cosmic string networks. Then, in Sec.~\ref{sec:sgwb}, we provide a brief overview of the computation of the SGWB genereated by cosmic string networks. In Sec.~\ref{sec:friction-network}, we describe the impact of friction on the evolution of a cosmic string network. In Sec.~\ref{sec:fricloop}, we study the impact of friction on the evolution of cosmic string loops numerically and derive an analytical approximation to describe the evolution of the length of loops during this era. Finally in Sec.~\ref{sec:sgwb-signature}, we characterise the signature of friction on the SGWB and study its dependence on the network and background plasma properties. We discuss the results and conclude in Sec.~\ref{sec:discussion}.

Throughout this paper, we will use natural units with $c=\hbar=1$, where $c$ is the speed of light in vacuum and $\hbar$ is the reduced Planck constant. Moreover, we will use cosmological parameters determined using Planck 2018 data~\cite{2020A&A...641A...6P}, where the values of the density parameters for radiation, matter and dark energy at the present time are respectively given by $\Omega_{\rm r} = 9.1476\cdot 10^{-5}$, $\Omega_{\rm m} = 0.308$, $\Omega_\Lambda=1-\Omega_{\rm r} - \Omega_{\rm m}$ and the Hubble constant is $H_0=2.13 \cdot h \cdot 10^{-33} \, \rm eV$, with $h=0.678$. 

\section{The evolution of cosmic string networks}\label{sec:network-evolution}

The evolution of cosmic string networks is determined mostly by four main physical processes. Strings have tension that force them to move and, in the absence of strong damping forces, this accelerates them to relativistic velocities. Cosmological expansion also has a significant impact throughout their evolution: it stretches long cosmic strings, rarefies the network density, and decelerates the strings. Also, as long strings move, they often collide with other strings or self-intersect. When this happens, the colliding strings exchange partners and reconnect, which may lead to the formation of closed loops of string. These loops detach from the network and evolve independently, thus resulting in energy loss. Finally, since cosmic string networks are expected to form deep in the early universe, strings inevitably interact with the particles of the background plasma in the early stages of their evolution. As a result,  strings also experience a frictional force, caused by these interactions, which plays a key role in the early stages of their evolution, when the universe was very dense.

\subsection{Cosmic String Evolution with Friction}

In the cosmological context, local cosmic strings may be regarded as infinitely-thin and featureless objects. They sweep, in their movement, a $1+1$-dimensional worldsheet in spacetime, represented by

\be
x^\mu=x^\mu(\sigma^0, \sigma^1)\,,
\ee
where $\sigma^0$ and $\sigma^1$ are, respectively, a timelike and a spacelike parameter of the worldsheet. Cosmic string dynamics may then be described by the Nambu-Goto action

\be 
S=-\mu\int \sqrt{-\gamma} d^2\sigma\,,
\ee 
where $\mu$ is the cosmic string mass per unit length (which, in this case, coincides with the cosmic string tension), $\gamma$ is the determinant of the worldsheet metric $\gamma_{ab}=g_{\mu\nu}x^\mu_{,a}x^\nu_{,b}$, with $a,b=0,1$, and $g_{\mu\nu}$ is the spacetime metric. This action, however, does not provide a full description of the evolution of cosmic strings in the early universe, as it does not take the interactions with particles of the background plasma into account. The frequent scattering of particles by cosmic strings leads, in fact, to a frictional force per unit length of the form~\cite{PhysRevD.43.1060}:

\be 
\mathbf{F}=-\frac{\mu}{l_f}\frac{\mathbf{v}}{\sqrt{1-v^2}}\,,
\ee 
where $\mathbf{v}$ is the string velocity. Here, we also introduced the friction length, $l_f$, that describes the characteristic lengthscale for which friction plays an important role in the dynamics of cosmic strings. For gauge strings, friction is mainly caused by Aharonov-Bohm scattering~\cite{Alford:1988sj} and this lengthscale assumes the form:
\begin{equation}
	\label{2.3}
	l_f = \frac{\mu}{\beta T^3}\,,
\end{equation}
where $T$ is the background temperature and $\beta$ is a parameter that depends on the number of particle species that interact with the string. Including the effect of friction, the equation of motion for a cosmic string is of the form~\cite{PhysRevD.43.1060}:

\be
{x^\nu_{,a}}^{;a}+\Gamma^\nu_{\sigma \mu}x^\sigma_{,a}x^{\mu,a}=\frac{1}{l_f} \left(U^\nu-x^\nu_{,a}x^{\sigma,a} U_\sigma\right)\,,
\ee
where $U^\nu$ is the 4-velocity of the background fluid and ${x^\nu_{,a}}^{;a}=\partial_a(\sqrt{-\gamma}\gamma^{ab}x^\nu_{,b})/\sqrt{-\gamma}$ is the covariant Laplacian.

In a Friedmann-Lemaitre-Robertson-Walker (FLRW) background, the line element is given by

\be 
ds^2=a(\eta)^2 (d\eta^2-d\mathbf{y}\cdot d\mathbf{y})\,,
\ee 
where $a$ is the cosmological scale factor, $d\eta=dt/a$ is the conformal time, $t$ is the physical time and $\mathbf{y}$ are cartesian coordinates, and the $4$-velocity of radiation is given by $U^\nu=(a^{-1},0,0,0)$. In this case, it is convenient to choose the temporal-transverse gauge, in which

\be 
\sigma^0=\eta \quad \mbox{and} \quad \dot{\mathbf{x}} \cdot {\mathbf{x}}'=0\,,
\ee 
where $x^\mu=(\eta,\mathbf{x})$ and dots and primes represent, respectively, derivatives with respect to $\eta$ and $\sigma\equiv \sigma^1$. The equations of motion for a cosmic string take then the form~\cite{Turok:1984db,PhysRevD.43.1060}:

\bq
\ddot{\mathbf{x}} +\left(2\frac{\dot{a}}{a}+\frac{a}{l_f}\right) \left(1-\dot{\mathbf{x}}^2\right) & = & \epsilon^{-1}\left(\epsilon^{-1} \mathbf{x}'\right)'\,,\label{eq:NGx}\\
\dot{\epsilon}+\left(2\frac{\dot{a}}{a}+\frac{a}{l_f}\right) \dot{\mathbf{x}}^2 \epsilon & = & 0 \,, \label{eq:NGepsilon}
\eq
where $\epsilon^2=\mathbf{x}'^2/(1-\dot{\mathbf{x}}^2)$.

\subsection{Velocity-dependent One-Scale model}

The cosmological evolution of a cosmic string network may be described, on sufficiently large scales, using the Velocity-dependent One-Scale (VOS) model~\cite{Martins:1996jp,Martins:2000cs}. This model provides a quantitative description of the cosmological evolution of cosmic string networks by following the evolution of two variables: the Root-Mean-Squared (RMS) velocity $\vv$ of the network, defined as

\be 
\vv^2\equiv \langle\dot{\mathbf{x}}^2\rangle= \frac{\int  \dot{\mathbf{x}}^2 \epsilon d\sigma}{\int \epsilon d\sigma}\,,
\label{eq:rms-def}
\ee 
and its characteristic lengthscale $L$, defined in terms of the energy density of long strings $\rho$  as

\be 
\rho=\mu/L^2\,,
\label{eq:L-def}
\ee
where $E=\mu a(\eta) \int \epsilon d\sigma$ is the energy of cosmic strings and $\rho \propto E a^{-3}$. For standard strings, without internal degrees of freedom, $L$ also measures roughly the average distance between long strings. The evolution equations for $\vv$ and $L$ may then be found by averaging Eqs.~(\ref{eq:NGx}) and~(\ref{eq:NGepsilon}) respectively\footnote{Note however that, in their derivation, it is assumed that $\langle\dot{\mathbf{x}}^4\rangle=\langle\dot{\mathbf{x}}^2\rangle^2=\vv^4$.} and by including a phenomenological term to account for the impact of collisions and interactions between cosmic strings. These interactions lead, as previously explained, to the production of cosmic string loops that detach from the Hubble flow and that are expected to decay and evaporate. This then results in an energy loss of the form~\cite{KIBBLE1985227}:

\be 
\left. \frac{d\rho}{dt}\right|_{loops}=\cc \vv \frac{\rho}{L}\,, \label{eq:loss}
\ee 
where $\cc=0.23\pm 0.04$~\cite{Martins:2000cs} is a phenomenological parameter that quantifies the efficiency of loop chopping. By using Eqs.~(\ref{eq:NGx})-(\ref{eq:loss}), one finds that the evolution of a cosmic string network on cosmological scales may be described by~\cite{Martins:1996jp,Martins:2000cs}:
\bq
	\frac{d\vv}{dt} & = &  (1 - \vv^2) \left[ \frac{k(\vv)}{L} - \vv \left( 2H + \frac{1}{l_f} \right) \right]\,,\label{eq:VOSv}\\
	2\frac{dL}{dt} & = &  2HL(1 + \vv^2) + \frac{L \vv^2}{l_f} + \cc\vv\,,\label{eq:VOSL}
\eq
where $k(\vv)$ is a phenomenological parameter that accounts, to some extent, for the effects of small-scale structure on long strings (here, we will use the \textit{ansatz} proposed in Ref.~\cite{Martins:2000cs}), and  $H=da/dt/a$ is the Hubble parameter. These equations enable us to describe quantitatively the cosmological evolution of cosmic string networks from early to late cosmological times. We briefly outline this evolution, with particular focus on the impact of friction in early cosmological times, in Sec.~\ref{sec:friction-network}.

\section{The Stochastic Gravitational Wave Background generated by cosmic string networks}\label{sec:sgwb}

After cosmic string loops are created, they detach from the Hubble flow and start to evolve under the effect of their tension. As a result, they oscillate periodically with relativistic velocities and they are then generally expected to decay by emitting GWs\footnote{The potential role of the emission of scalar and gauge radiation is currently still a matter of debate~\cite{Hindmarsh:2017qff,Hindmarsh:2021mnl,Blanco-Pillado:2022rad,Blanco-Pillado:2023sap}. Abelian-Higgs simulations indicate that this radiation may lead to the fast decay of the majority of loops~\cite{Hindmarsh:2017qff} and that less than $10\%$ of the loops produced would decay by emitting GWs~\cite{Hindmarsh:2021mnl}. It is argued in~\cite{Blanco-Pillado:2022rad,Blanco-Pillado:2023sap}, however, that this may be a transient phenomenon.}. The superimposition of the GW bursts generated by the copious amounts of loops that are created throughout the evolution of the cosmic string network gives rise to a Stochastic Gravitational Wave Background (SGWB)~\cite{Vilenkin:1981bx,hogan,accetta}. This SGWB is generally characterized by the spectral energy density of GWs,

\begin{equation}
    {\Omega}_{\rm gw}(f)=\frac{1}{\rho_c}\frac{d{\rho}_{\rm gw}}{d\log f}\,,
\end{equation}
where ${\rho}_{\rm gw}$ is the energy density of gravitational radiation, $\rho_c=3H_0^2/(8\pi G)$ is the critical density of the universe at the present time $t_0$ (for the remainder of this paper, the subscript `$0$' is used to refer to the value of the corresponding quantity at $t=t_0$).

The frequency of the GWs emitted by cosmic string loops is determined by harmonics of their length $\ell$ at the time of emission $t$. They then arrive to an observer at the present time $t_0$ with a frequency
\begin{equation}
	\label{3.1}
    f_j = \frac{2j}{\ell} \frac{a(t)}{a_0}\,,
\end{equation}
where $j$ is the harmonic mode of emission and the subscript `$j$' is used to label the contribution of the $j$-th mode of emission to the corresponding quantity. The spectrum of emission of cosmic string loops follows roughly a power law~\cite{Burden:1985md,Garfinkle:1988yi,Allen:1994bs,Damour:2000wa,Blanco-Pillado:2015ana} of the form

\begin{equation}
    \frac{dE_j}{dt}=\Gamma_j G\mu^2\,, \quad \mbox{with}\quad \Gamma_j=\frac{\Gamma}{\zeta(q)}j^{-q}\,,
    \label{eq:powerspectrum}
\end{equation}
where $E=\mu\ell$ is the energy of the loops, $\zeta(q)$ is the Riemann Zeta function and $\Gamma\sim 50$~\cite{Vachaspati:1984gt,Burden:1985md,Scherrer:1990pj,Quashnock:1990wv,Allen:1994bs,Blanco-Pillado:2015ana} is  the GW emission efficiency~\footnote{Note that these studies only apply to cosmic strings without internal degrees of freedom. If strings carry currents, for instance, $\Gamma$ should be suppressed and the spectrum of emission may be altered~\cite{Rybak:2022sbo}.}. The spectral index $q$ depends on the small-scale structure present in the cosmic string loops. Loops are generally expected to have points that move instantaneously at the speed of light, known as cusps, that give rise to a spectrum characterized by $q=4/3$. Moreover, collisions between strings lead to discontinuities in the string that travel along the string at the speed of light. These kinks generate a spectrum characterized by $q=5/3$ and, when two kinks collide, by $q=2$.

The amplitude of the SGWB generated by cosmic string loops is then given by (see e.g.~\cite{Vilenkin:2000jqa}):

\begin{equation}
   {\Omega}_{\rm gw}(f)=\sum_{j=1}^{+\infty} \Gamma_j {\Omega}_{\rm gw}^j(f)\,,
\end{equation}
where

\begin{align}
	\label{SGWB}
    {\Omega}_{\rm gw}^j(f) =\frac{16 \pi}{3f}\left(\frac{G\mu}{H_0}\right)^2 \int_{t_{i}}^{t_0} j\, n\left(\ell_j(t), t\right) \left( \frac{a(t)}{a_0} \right)^5dt \nonumber \\
\end{align}
is the contribution of the $j$-th harmonic mode of emission and $\ell_j=2ja(t)/(f a_0)$ is the length of loops that radiate, in the $j$-th harmonic, GWs that have a frequency $f$ at the present time. Moreover, $t_i$ is the time in which significant emission of gravitational radiation by cosmic string loops starts, which is often assumed to be the end of the friction-dominated era. In this paper, we will study the validity of this assumption and examine whether there are signatures of friction on the SGWB. We will then generally assume that $t_i$ coincides with the time of creation of the network $t_c$. We will also consider the contribution of the fundamental mode of emission (and drop the superscript `$1$' that denotes this mode). Note however that it is straightforward to compute the contribution of any mode of emission once that of the fundamental mode is known:

\begin{equation}
    {\Omega}_{\rm gw}^j(jf)={\Omega}_{\rm gw}^1(f)\,,
\end{equation}
and thus considering $j=1$ is sufficient to fully characterize the SGWB.

The crucial ingredient to compute the SGWB generated by cosmic string loops is then the loop distribution function,  $n(\ell,t)d\ell$, which describes the number density of string loops with lengths between $\ell$ and $\ell+d\ell$ that exist at the time $t$:
\begin{equation}
	\label{numberdensity}
n(\ell,t) = \int_{t_i}^{t} dt_b f(\ell_b, t_b) \left( \frac{a(t_b)}{a(t)} \right)^3\,,
\end{equation}
where $\ell_b$ is the length of the loop at the time of birth $t_b$. Here $f(\ell, t)d\ell$ is the loop production function, which represents the number density of loops with lengths between $\ell$ and $\ell + d\ell$ produced per unit time. The amount of energy that is lost as a result of loop production may be inferred from the large-scale dynamics of the cosmic string network using Eq.~(\ref{eq:loss}). If one assumes that the length of loops at the moment of creation is proportional to the characteristic length of the long string network at that instant of time,
\begin{equation}
	\label{3.4}
\ell(t_b) = \alpha L(t_b)\, ,
\end{equation}
where $\alpha < 1$ is a constant loop-size parameter, we then have that~\cite{Sousa:2013aaa,auclair,Sousa:2020sxs}:
\begin{equation}
	\label{f}
	f(\ell,t) = \frac{\cc}{\sqrt{2} \ell} \frac{\vv(t)}{L(t)^3} \delta(\ell - \alpha L)\,,
\end{equation}
where the factor of $\sqrt{2}$ was introduced to account for the effect of the peculiar velocities of loops~\cite{Vilenkin:2000jqa}. Note that, although the assumption that all loops are created with the same length seems rather strong, this form of the loop production function may be used to construct the loop production function for any distribution of loop lengths at birth~\cite{Sanidas:2012ee,Sousa:2020sxs}. Moreover, the impact of the length of loops following a distribution at the moment of creation on the SGWB may also, to some extent, be described by including a factor $\mathcal{F}<1$ in Eq.~(\ref{3.4})~\cite{auclair,Sousa:2020sxs}. Such a factor could also account for the possibility that not all loops decay by emitting GWs. Here, for simplicity, we take $\mathcal{F}=1$. We then have that the loop distribution function is given by~\cite{Sousa:2013aaa}:
\bq
    \label{number_dencity_2}
    n(\ell,t) = \frac{dt_b}{dl_b} \frac{\cc \vv\left( t_b \right)}{\sqrt{2} \alpha L\left( t_b \right)^4} 
    \left( \frac{a \left( t_b \right)}{a \left( t \right)} \right)^3 \,.
\eq

\section{Impact of friction on the evolution of cosmic string networks}\label{sec:friction-network}

Cosmic string networks are expected to form in the early universe at a critical temperature $T_c$ that determines their mass per unit length: $\mu \sim T_c^2$~\cite{Kibble}. The time of string formation is then roughly given by:
\be
    \label{t_c}
    t_c = \frac{1}{\chi(t_c)} \frac{t_{pl}}{G \mu}, \quad\mbox{with}\quad \chi(t) = 4 \pi \left( \frac{\pi g_{*}(t)}{45} \right)^{1/2}\,,
\ee
where $t_{pl}=G^{1/2}$ is the Planck time and $g_{*}(t)$ is the effective number of massless degrees of freedom (see e.g.~\cite{Kolb:1990vq}). The friction epoch happens, in general, long before the first change $g_{*}(t)$ predicted by the Standard Model of Particle Physics (which is assumed throughout our calculations). For the remainder of this paper, we will then assume that $\chi(t)=\chi(t_c)\equiv \chi$ throughout the friction era.

Cosmic string networks, however, are generally expected to survive throughout cosmic history and, depending on what damping mechanism dominates their dynamics, they can go through different scaling regimes in their evolution. 
In the early universe, interactions between cosmic strings and the particles of the surrounding plasma should be quite frequent and, as a result, their dynamics should be dominated by friction. In this case, string movement is heavily damped and they are expected to move with non-relativistic velocities and, as can be inferred from Eq.~(\ref{eq:VOSv}), their velocity is roughly given by $\vv \simeq k_c l_f/L$, where $k_c\equiv k(0)=2\sqrt{2}/\pi$. During this friction-dominated era, two different scaling regimes may emerge. The first is the \textit{Stretching Regime}, during which strings are frozen in comoving coordinates and stretched by expansion:

\be 
L \propto a\quad\mbox{and}\quad \vv \propto a^2\,.
\ee 
Assuming that the friction-dominated epoch occurs deep in the radiation era (which, as we shall see, should in general be the case) we should have that, during this regime,

\begin{equation}
	\label{stretching}
	L = L_c \left( \frac{t}{t_c} \right)^{1/2}, \quad \mbox{and} \quad	\vv = \vv_{c} \left( \frac{t}{t_c} \right)\,,
\end{equation}
where $L_c$ and $\vv_{c}=k_c l_f(t_c)/L_c$ are the initial characteristic lengthscale and the RMS velocity, respectively. Note that, in general, we should have that $l_f (t_c)< L_c<t_c$~\cite{Martins:1996jp}.

During the Stretching regime, the movement of strings is so damped by friction that there are almost no interactions between strings. However, as the universe expands and cools, the friction lengthscale $l_f$ grows much faster than the characteristic length $L$ (cf. Eq.~(\ref{2.3}) and~(\ref{stretching})) and interactions become increasingly relevant. As the Hubble damping and friction terms in Eq.~(\ref{eq:VOSL}) become comparable (i.e., when $l_f/L\sim H L$), the \textit{Kibble Regime}, characterized by

\begin{equation}
	\label{2.5}
	L \propto \left(\frac{l_f}{H}\right)^{1/2} \quad \mbox{and} \quad 
	\vv \propto \left (l_f H\right)^{1/2}\,,
\end{equation}
emerges. In this regime, a considerable amount of energy is lost as a result of interactions, since $HL \sim \cc \vv$, which explains why the network evolves differently. Loop production is not negligible during the Kibble regime. As a matter of fact, as we will show, it is quite significant, since networks are expected to be quite dense during this stage. 

More precisely, the Kibble regime should emerge when the following condition is satisfied

\be
    \label{Kibble_condition}
    AHL = \frac{l_f}{2L} k_c(k_c+\cc)\,,
\ee
where $A$ is a constant of order unity that was introduced to obtain a better fit to the numerical evolution. During the radiation era, we may express the friction lengthscale as~\cite{Martins:1996jp}:
\be
    \label{friction_length_2}
    l_f = \frac{1}{\theta}\frac{t^{3/2}}{t_c^{1/2}}\,, \quad \mbox{with}\quad
    \theta = \frac{\beta}{\chi\sqrt{G\mu}}\,.
\ee
We then have that, during the Kibble regime, 
\bq
    \frac{L}{t_c}  & = & \left( \frac{k_c(k_c+\cc)}{\theta A} \right)^{1/2} \left( \frac{t}{t_c} \right)^{5/4}\,,\label{L_Kibble}\\
    \vv & = & k_c \left( \frac{A}{\theta k_c (k_c+\cc)} \right)^{1/2} \left( \frac{t}{t_c} \right)^{1/4}\,.\label{v_Kibble}
\eq
These equations may be confronted with the full numerical evolution during the Kibble regime --- described by Eqs.~(\ref{eq:VOSv}) and~(\ref{eq:VOSL}) --- to determine the value of the constant $A$ in Eq.~(\ref{Kibble_condition}). Our analysis indicates that $A=3/2$ provides an excellent fit for the relevant range of initial conditions and for different values of $\beta$.

\begin{figure}
	\begin{center}
		\centering
		\includegraphics[width=\linewidth]{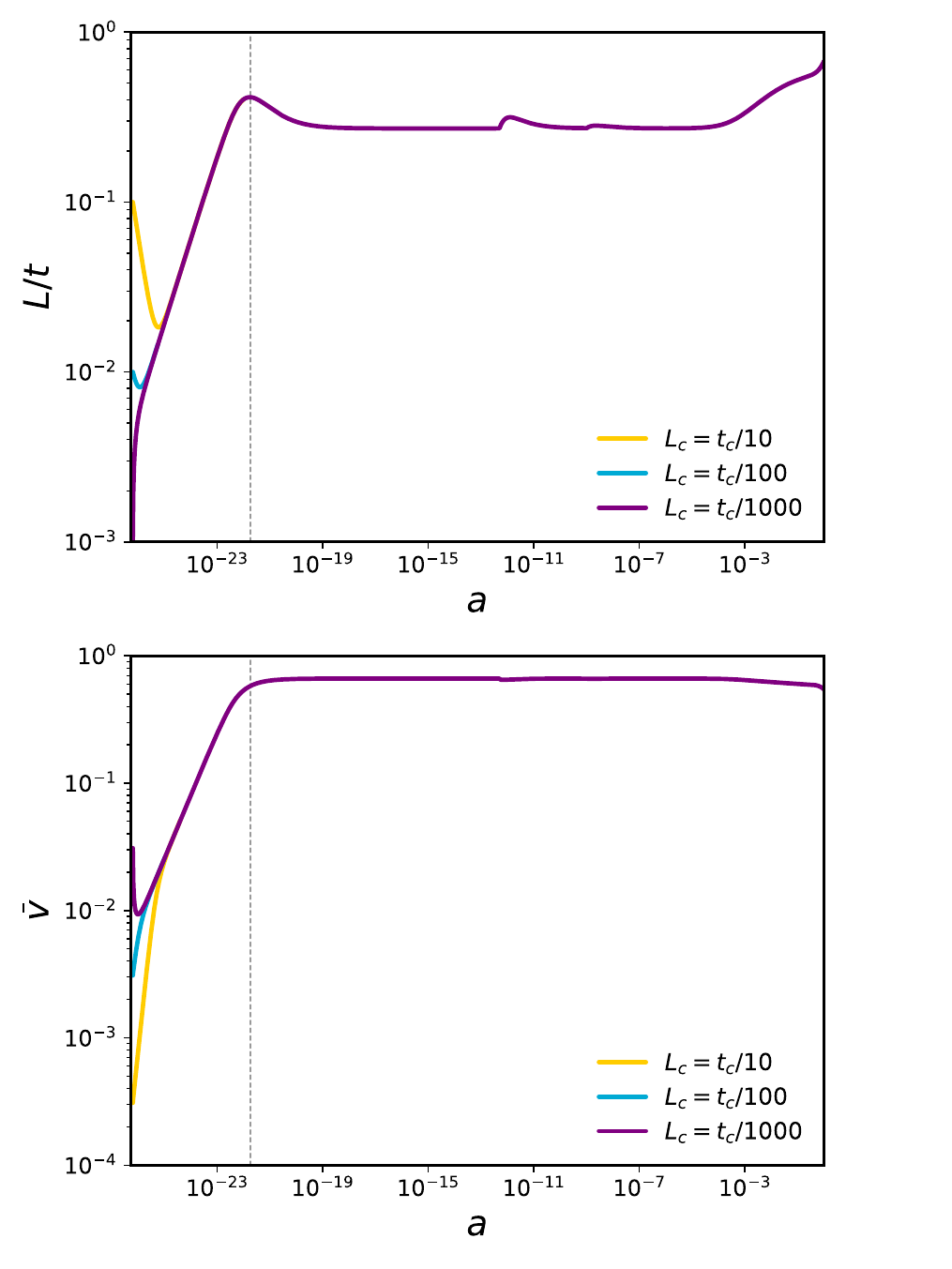}
		\caption{Evolution of a cosmic string network with friction for $G\mu = 10^{-12}$. The top panel displays the characteristic length divided by the physical time and we plot the RMS velocity on the bottom panel. The solid lines of different colours correspond to different initial conditions. The vertical dashed line represents the value of the scale factor at the end of the friction dominated epoch, $a_f$.}
		\label{fig:Full_evolution}
	\end{center}
\end{figure}

To estimate the time in which the network enters the Kibble regime, we may assume that the transition between the Stretching and Kibble regimes happens suddenly at a time $t_k$, corresponding to the time in which the condition in Eq.~(\ref{Kibble_condition}) is first satisfied. Assuming that the network is in the Stretching regime characterized by Eq.~(\ref{stretching}) for $t \leqslant t_k$, we have that

\be
    \label{t_k}
    t_k = \left( \frac{3\theta}{2k_c (k_c+\cc)} \right)^{\frac{2}{3}} \left( \frac{L_c}{t_c} \right)^{\frac{4}{3}} t_c\,.
\ee
This expression shows that the time of emergence of the Kibble regime depends on the initial conditions and thus so does the duration of the Stretching regime. As a matter of fact, the network only goes through the Stretching regime if the conditions of existence of the Kibble regime are not met initially (or, in other words, provided that the initial characteristic length is not significantly smaller than the horizon). However, the cosmic string network necessarily goes through the Kibble regime in the early universe, either following the Stretching regime or right after creation if its initial density is large enough. This may be seen clearly in Fig.~\ref{fig:Full_evolution}, where the evolution of $L/t$ and $\vv$ are plotted for different initial conditions. Therein, one may see that, if the initial characteristic length of the network is smaller, the Kibble regime starts earlier and the duration of the Stretching regime decreases.

The friction-dominated era of cosmic string evolution is, however, necessarily transient. As the friction lengthscale continues to increase rapidly, it will eventually become larger than the characteristic length of the network and the impact of friction on the dynamics of cosmic string networks will become negligible. Assuming as before that this transition to the frictionless epoch happens suddenly at a time $t_f$ in which the Hubble damping term $2H$ becomes comparable to the friction damping term $l_f^{-1}$, we have that:
\be
    \label{t_f}
    t_f = \theta^2 t_c\,.
\ee
Using this one may show that, unless cosmic string tension is very small, string dynamics should become frictionless before any variation of $\chi(t)$ (according to the Standard Model of Particle Physics) and well before the radiation-matter transition, which justifies our assumption that friction happens deep in the radiation era.

After $t_f$, when friction becomes negligible to cosmic string dynamics, the movement of strings is no longer so heavily damped and the network evolves towards a Linear scaling regime characterized by:
\begin{equation}
	\label{2.6}
	L = \xi t, \;\; \frac{d\vv}{dt} = 0 \,,
\end{equation}
with
\be
    \label{linear}
    \xi^2 = \frac{k(k+\cc)}{4\nu(1-\nu)}, \;\; \vv^2 = \frac{k}{k+\cc} \frac{1-\nu}{\nu}\,.
\ee
This regime is an attractor solution of the VOS equations when $a\propto t^{\nu}$, with $0<\nu<1$, and its existence has been verified in numerical simulations. In this regime, cosmic strings reach relativistic velocities and, as a result, interactions between strings and loop production are quite frequent. As Fig.~\ref{fig:Full_evolution} illustrates, the evolution of the cosmic string network during the frictionless epoch, for $t>t_f$, does not depend on the initial conditions. Note that the cosmic string network is temporarily ``knocked out'' of the Linear Scaling regime as a result of the decrease in the effective number of massless degrees of freedom, as the universe cools. This is the cause of the small bumps seen, during the radiation-era Linear scaling regime, in Fig.~\ref{fig:Full_evolution} (these are more evident in the plot for $L$ due to the scale of the plot, but also show up in the plot for $\bar{v}$ in the form of a small temporary decrease).  Also, later the universe transitions to the matter-dominated era and the network starts to evolve towards a new Linear Scaling regime. However, this regime is not reached before the present time due to the emergence of dark energy. Once this happens, cosmic string movement is heavily damped by the fast expansion, which leads to a new stretching regime of the form $L\propto a$ and ${\bar v}\to 0$. In this case, the Stretching regime is sustainable because the accelerated expansion ensures that the network remains frozen, with non-relativistic velocities.

\section{Impact of friction on the evolution of cosmic string loops}\label{sec:fricloop}

To include the contribution of the loops created during the friction-dominated regime in the computation of the SGWB, we need to start by studying the evolution of a cosmic string loop in the presence of friction. Although this has been studied previously in~\cite{PhysRevD.48.2502}, therein the impact on gravitational wave emission --- which is crucial for our study --- was not considered. Here, we follow a similar approach and consider a circular loop with a coordinate radius $q$:
\begin{equation}
\label{eq:loop-solution}
\mathbf{x} = q(\eta) (\sin\phi, \cos\phi, 0)\,,
\end{equation}
and choose the spacelike worldsheet parameter to be $\phi$. In a FLRW universe, the physical length of the loop $\ell$ --- defined such that $E=\mu \ell$, where $E$ is the energy of the loop --- is given by
\begin{equation}
	\label{length}
\ell = 2 \pi \gamma |q| a\,,
\end{equation}
where $v=\dot{q}$ is the velocity of the loop and $\gamma=(1-v^2)^{-1/2}$.

Cosmic string loops are usually assumed to emit GWs at a roughly constant rate given by
\begin{equation}
    \left. \frac{d\ell}{dt}\right|_{\rm gw}=\Gamma G\mu\,,
    \label{eq:gwemission-av}
\end{equation}
with $\Gamma=50$ (as may be seen from Eq.~(\ref{eq:powerspectrum}), by summing over all the harmonic modes of emission). However, more generally, the energy loss caused by the emission of gravitational radiation, which may be estimated using the quadrupole formula, may be written in the following form~\cite{Martins:1996jp}:

\begin{equation}
	\label{dldt_back_full}
	\frac{d\ell}{dt}\bigg|_{GW} = -\Gamma' G \mu v^6\,,
\end{equation}
where $\Gamma'$ is a parameter evaluating the efficiency of GW emission. The approximation in Eq.~(\ref{eq:gwemission-av}), in fact, only describes the decrease of the length of loops caused by the emission of gravitational radiation in an averaged sense. Cosmic string loops, in the absence of friction and provided that expansion has a negligible impact, oscillate periodically with an average squared velocity of $\langle v^2 \rangle =1/2$. In these oscillations, they reach ultra-relativistic velocities periodically and they should, in fact, emit GWs dominantly in the stages of the oscillation in which their velocities are significant. In general, for these freely oscillating loops, assuming a constant rate of GW emission provides an adequate description of their evolution on time-scales much larger than their oscillation period. Here, since we are including the impact of friction --- which damps the movement of loops and may significantly delay their oscillations~\cite{PhysRevD.48.2502} ---, we use the form in Eq.~(\ref{dldt_back_full}) to account for the possibility that the rate of GW emission can no longer be assumed to be constant. As proposed in~\cite{Martins:1996jp}, for consistency, given the fact that for freely-oscillating loops $\langle v^6 \rangle =5/16$, we take $\Gamma'=16/5 \cdot \Gamma=160$ \footnote{In Ref.~\cite{Martins:1996jp}, the authors take $\Gamma'=8\Gamma$ since, by mistake, they assume that $\langle v^6\rangle=\langle v^2\rangle ^3$.}.

By introducing Eq.~(\ref{eq:loop-solution}) into Eqs.~(\ref{eq:NGx}) and~(\ref{eq:NGepsilon}) and using Eqs.~(\ref{length}) and (\ref{dldt_back_full}), one finds that the equation of motion for $q$ is given by:
\begin{equation}
	\label{dqdt}
	\ddot q = (1 - \dot q^2) \left( - 2Ha\dot q - \frac{a \dot q}{l_f} - \frac{1}{q} -
	\frac{1}{2 \pi |q|} \sqrt{1 - \dot{q}^2} \Gamma' G \mu \dot{q}^5 \right)\,.
\end{equation}
To study the impact of friction on the evolution of cosmic string loops and on their GW emission, we solved this equation numerically, with initial conditions of the form
\begin{equation}
	\label{initialconditions}
	q_b = \frac{\alpha L(t_b)}{2 \pi a_b}\quad\mbox{and} \quad
    v_b = \dot q_b = 0\,,
\end{equation}
for a wide range of the loop-size parameter $\alpha$ and of the birth time of loops $t_b$. This choice of initial conditions is natural since it is well known that a free circular loop will oscillate with amplitude $q_b$ having vanishing velocity when the loop radius is at a maximum. This extensive study was done with the objective of determining which physical processes play an important role in the evolution of subhorizon loops --- the main contributors to the SGWB --- and to aid us in the development of an analytical approximation to the evolution of $\ell$.

Our results have shown that for very large loops (with lengths larger or comparable to the horizon), the expansion of the background stretches loops significantly thereby increasing their energy --- and particularly so in the presence of friction as was also found in~\cite{Vilenkin:1984ib,Garriga} --- but small enough loops do not feel its effects even when friction is present. Here, since we are considering sub-horizon loops, we will then neglect the expansion term in Eq.~(\ref{dqdt}), as is usually done in the absence of friction. Notice however that the impact of expansion on the evolution of the friction lengthscale cannot be neglected, as it is determined by the density of the background plasma which is strongly dependent on the expansion.

We have also verified numerically that, for sub-horizon loops, even deep in the friction-dominated regime, we have $\langle v^2\rangle\approx 1/2$ as for free loops evolving in the absence of expansion, friction and GW emission. One may then conclude that, during the friction era, sub-horizon loops also emit GWs at a roughly constant rate, with $\Gamma = 50$, and, in fact, we have verified numerically that the average energy lost in the form of gravitational radiation in a period of oscillation computed using Eq.~(\ref{dldt_back_full}) is well described by the simpler expression in Eq.~(\ref{eq:gwemission-av}).

Finally, our results also show that the dynamics of any loop will eventually become frictionless. This may be explained by the fact that $\ell$ decreases over time as a result of the energy loss caused by friction and by the emission of gravitational radiation, while the friction lengthscale increases (see Eq.~(\ref{2.3})). Thus as time goes by, the impact of friction becomes smaller and smaller until loop dynamics becomes frictionless and, once this stage is reached, the rest of the energy of the loop will be radiated in the form of GWs. Notice that the later in the friction era a loop is created, the sooner this frictionless regime will be reached and then an increasingly larger fraction of the energy of loops will be converted into GWs as time ellapses. For instance, for a loop with $G \mu = 10^{-12}$ and $\alpha = 0.1$ created deep in the friction era ($t_b = 10^{-5} \cdot t_f$) only about $ 0.2 \, \%$ of its energy will remain by the time this frictionless stage is reached. However, closer to the end of the friction epoch (for $t_b = 10^{-2} \cdot t_f$), without the inclusion of GW emission, the loop still has about $56 \, \%$ of its initial energy at this stage. This shows that, in fact, there is GW emission throughout the friction regime.

To calculate the number density of loops with a length $\ell$ at a time $t$, we need to find the time of formation of these loops, which means that we need to follow their evolution since the time of birth. However, for the physically relevant values $\alpha$ and $G\mu$, loops lose energy very slowly during most of the evolution and undergo a very large number of oscillations before evaporating. As a result, the numerical computation of their evolution is time costly. It is then convenient to find an analytical approximation to describe the evolution of cosmic string loops, including the effects of friction and GW emission, to make the computation of the SGWB more efficient. Neglecting, as suggested by our analysis, the Hubble damping term in Eq~(\ref{dqdt}) and assuming that GW emission happens with the constant rate given by Eq.~(\ref{eq:gwemission-av}), the evolution of $\ell$ should then be well described by
\begin{equation}
	\label{dldt}
	\frac{d\ell}{dt} = -\frac{\ell v^2}{l_f} - \Gamma G \mu\,,
\end{equation}
as may be found by combining Eqs.~(\ref{length}) and~(\ref{dqdt}) under these assumptions. Let us start by considering the case of a loop evolving only under the effect of friction and curvature and drop the GW emission term in Eq.~(\ref{dldt}). To find an analytical approximation to describe the evolution of $\ell$ on time scales much larger than a period of oscillation, we may then average this expression over one period of oscillation and assume that $\langle v^2\rangle =1/2$. We then find that
\be 
	\label{lfr}
	\ell = \ell_b exp\left[ t_f^{1/2}\left( t^{-1/2} - t_b^{-1/2} \right) \right]\,.
\ee
This expression is equivalent to the expression found in \cite{PhysRevD.48.2502} and also supports the observation that the effect of friction will eventually become negligible. Indeed, when $t\to\infty$, we see that $\ell$ does not vanish. The loop then does not evaporate completely as a result of friction and at least a fraction of $\exp\left(-t_f^{1/2}/t_b^{1/2}\right)$ of the initial length of the loop should be converted into GWs.

\begin{figure}
	\begin{center}
		\centering
		\includegraphics[width=\linewidth]{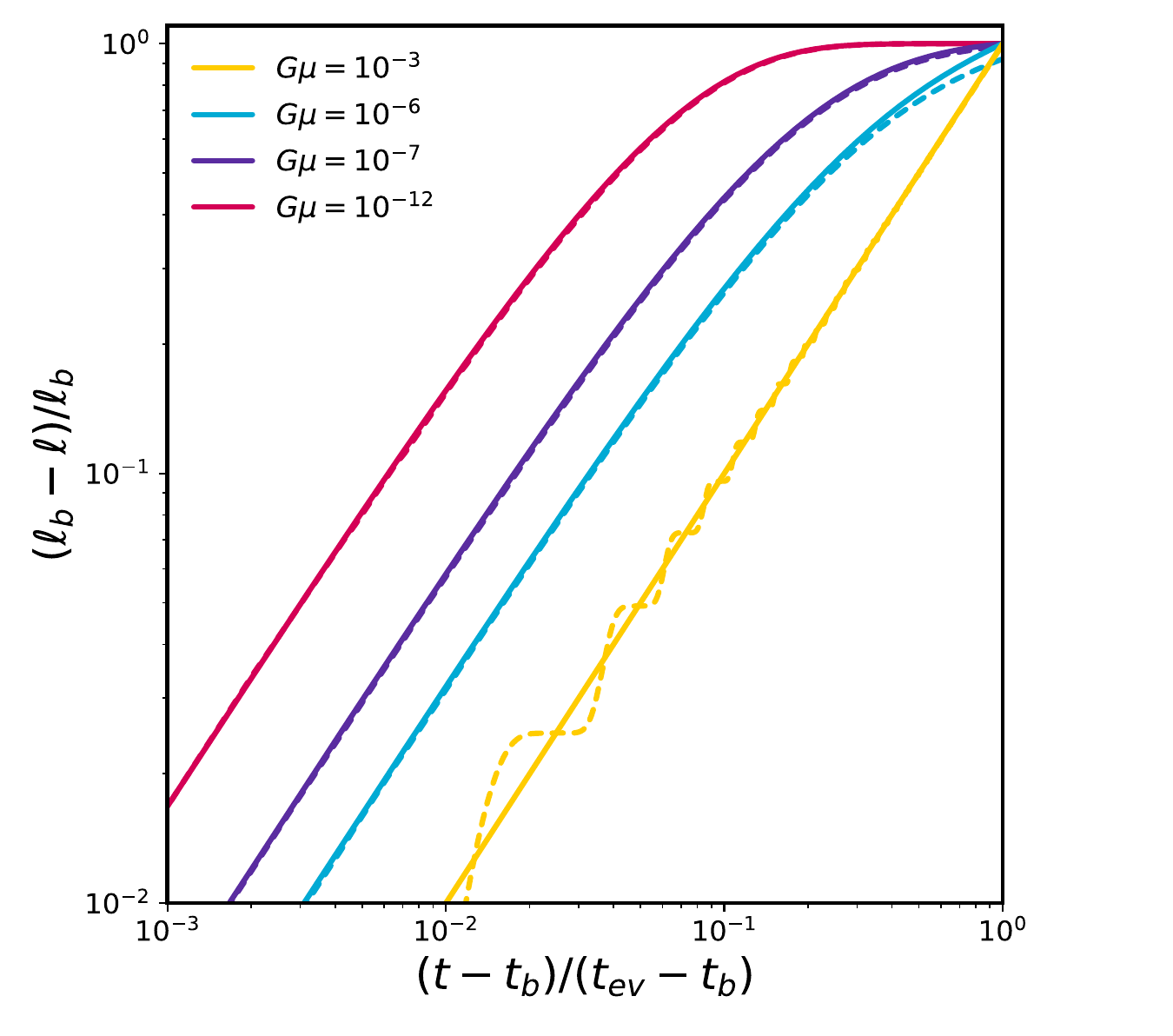}
		\caption{Evolution of loops with $\alpha = 10^{-3}$ and $t_b = t_k$ in the presence of friction for different values of cosmic string tension $G \mu$. Here, the solid lines correspond to the evolution predicted by the analytical approximation in Eq.~(\ref{l}), while the dashed lines represent that computed numerically by solving Eq.~(\ref{dqdt}).}
		\label{fig:loops_evolution}
	\end{center}
\end{figure}

Introducing the energy loss caused by the emission of gravitational radiation in Eq.~(\ref{lfr}), we have:
\be
	\label{l}
	\ell = \ell_b exp\left[t_b^{1/2} \left( t^{-1/2} - t_b^{-1/2} \right) \right] - \Gamma G \mu \left( t - t_b \right)\,.
\ee
We have compared this approximation extensively to the full numerical evolution described by Eq.~(\ref{dqdt}), including the curvature, friction and GW radiation terms in their complete form. Our results show that this approximation provides a good description for the evolution of the length of the loops produced during the Kibble regime --- which is enough since the contribution of the loops created during the Stretching regime to the SGWB is negligible (as we shall see later). In Fig.~\ref{fig:loops_evolution}, we display some examples of the evolution of the length of the loops throughout their lifetime (computed numerically) alongside the analytical approximation derived here. We plot $(\ell_b - \ell)/\ell_b$ as a function of $(t-t_b)/(t_{ev}-t_b)$, where $t_{ev}$ denotes the moment of evaporation of the loop, since this choice allows us to reflect all stages of their evolution and to plot several cases on the same figure. We display loops with $\alpha = 10^{-3}$ for four different values of cosmic string tension so that, as tension decreases, the impact of friction becomes stronger. Since the period of oscillation of loops is proportional to their length and, for small values of $G\mu$, they lose energy very slowly, the numerical study of the later stages of their evolution is computationally costly. 
For this reason, for illustration purposes, we opted to display the evolution of loops with unphysically large tensions too ($G \mu = 10^{-3}$ and $G \mu = 10^{-6}$) as this allows us to follow the evolution of loops numerically until $t_{ev}$ in the case of strong GW emission. The slope of the tangent vector of each curve at the time of evaporation characterises the fraction of energy which was lost in the form of GWs, reaching its maximum when GW emission is at its strongest for the largest $G \mu$. This figure clearly shows that our approximation provides an excellent description of the evolution of a loop when friction is strong (the beginning of the evolution for the lowest tension) --- which demonstrates the validity of our assumption that $\langle v^2\rangle\approx 1/2$ --- and when its effects are negligible and the evolution is determined by the emission of GWs instead (high tensions). When the impacts of friction and GW emission becomes comparable, however, our approximation predicts a slightly faster decay of the loops, which results in an underestimation of the time of evaporation of the loop. Note however that this does not have a significant impact on the quality of the results we will derive since assuming a faster decay leads, in general, to smaller contributions to the SGWB. These results are then conservative and may then be seen as a lower bound on the amplitude of the spectrum.

\section{The signature of friction on the SGWB}\label{sec:sgwb-signature}

It is usually assumed in the literature that the loops created during the friction-dominated epoch do not provide a significant contribution to the SGWB, since their motion would be effectively damped by friction. However, during this era, the network is very dense, which makes intersections and, consequently, the creation of loops more probable. Also, as we have seen, although friction is effective in the initial stages, the evolution of loops eventually becomes frictionless and, from this point on, their energy is lost only in the form of gravitational radiation. The combination of these two effects may result in a significant contribution to the SGWB.

In this section, we fully characterize the signature of friction on the SGWB, which involves a detailed computation of the loop distribution function during this era. Notice that the impact of friction on $n(\ell,t)$ is twofold. On the one hand, friction, by significantly affecting the large-scale evolution of the cosmic string network --- so much so that it can no longer be assumed to evolve in a linear scaling regime --- affects the loop production function. Here, we use the method introduced in~\cite{Sousa:2013aaa} (and summarized in Sec.~\ref{sec:sgwb}) and solve the VOS Eqs.~(\ref{eq:VOSv}) and~(\ref{eq:VOSL}), coupled with the Friedmann equation, to fully characterize $f(\ell,t)$ throughout the evolution of the cosmic string network. Our results show that indeed there is an intensive loop production during the friction epoch. For instance, for $G \mu = 10^{-12}$ the number of loops produced at the beginning of the Kibble regime is $400$ times larger than what one would have if the network was in a linear scaling regime. On the other hand, $n(\ell,t)$ is also affected by the change in the rate of energy loss of the loops caused by this additional decay mechanism, which may reduce the lifetime of the loops significantly. Here, we use the analytical approximation in Eq.~(\ref{l}) to find the Jacobian $dt_b/d\ell_b$ needed to fully characterize the loop distribution function in Eq.~(\ref{number_dencity_2}):
\be
    \label{dl_b/dt_b}
    \frac{dt_b}{d\ell_b} = \left( \alpha \frac{dL}{dt} \bigg|_{t_b} + \frac{\alpha L(t_b)}{2 t_b^{3/2}}t_f^{1/2}  + \Gamma G \mu \right)^{-1}
\ee
and to find the time of birth $t_b$ of the loops that contribute to the given frequency.

\subsection{The signature of friction}

\begin{figure}
		\centering
		\includegraphics[width=\linewidth]{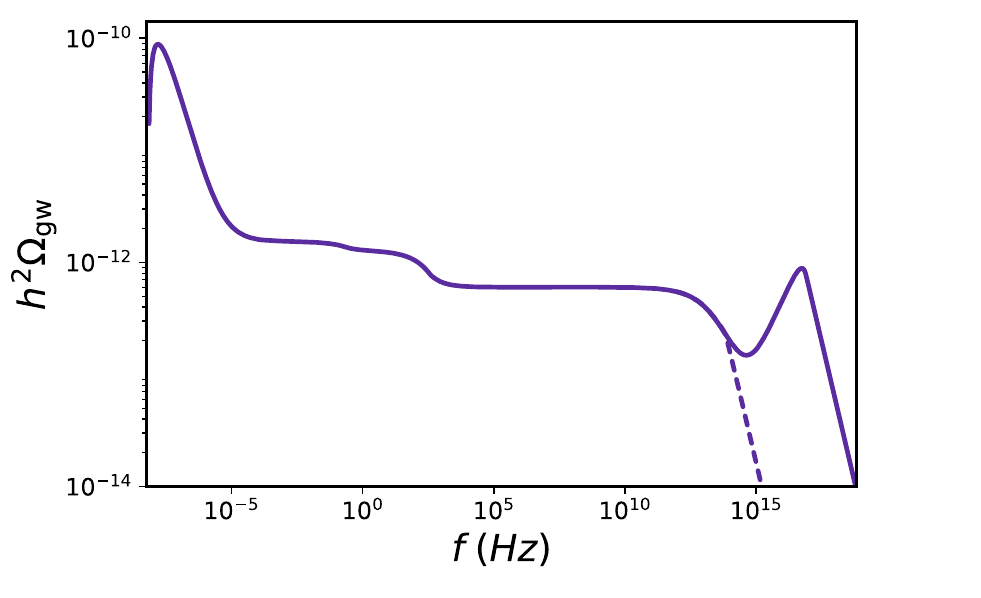}
		\caption{The Stochastic Gravitational Wave Background generated throughout the full evolution of a cosmic string network with $G \mu = 10^{-10}$, $\alpha = 10^{-9}$, $\beta = 1$, $L_c=t_c$. Here, the solid line represents the spectra including the contribution of the loops created during the friction-dominated era of cosmic string network evolution, while the dashed line corresponds to the spectra one obtains by assuming that significant GW emission by cosmic string loops starts only at the beginning of the frictionless regime.}
		\label{fig:SGWBfull}
\end{figure} 

\begin{figure}
	\begin{center}
		\centering
		\includegraphics[width=\linewidth]{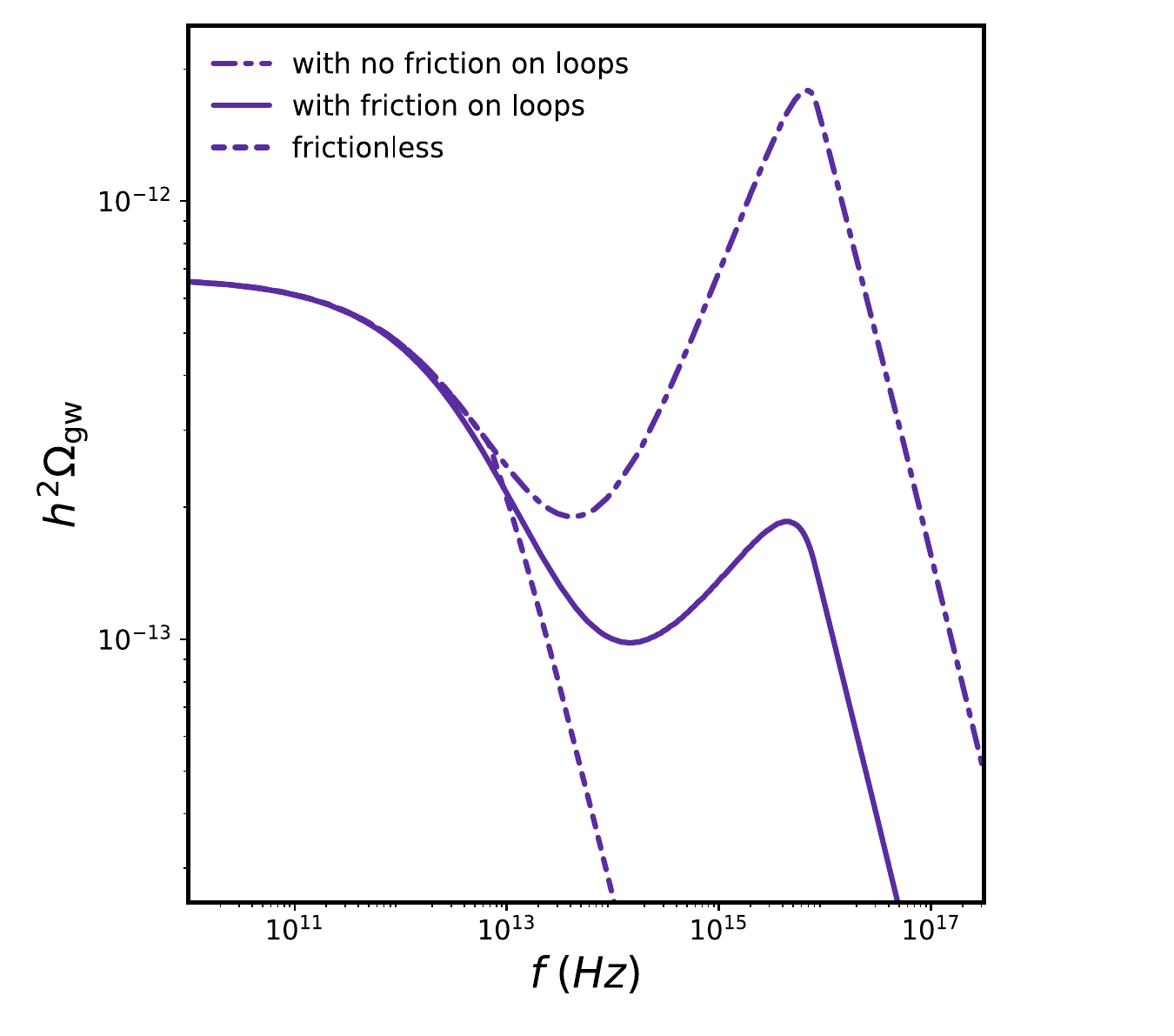}
		\caption{The impact of the inclusion of friction in the evolution of cosmic string loops on the signature of friction. The solid line represents the full SGWB generated by a cosmic string network during the friction era for $G\mu = 10^{-10}$, $\alpha = 10^{-8}$, $\beta = 1$, $L_c=t_c$. The dashed line corresponds to the SGWB one would obtain by completely neglecting the contribution of the loops created during the friction-dominated era of cosmic string network evolution, while the dash-dotted line represents that one would obtain if the impact of friction on cosmic string loops was not included.}
		\label{fig:nofriction}
	\end{center}
\end{figure}

We now have all the ingredients to characterize the signature of friction on the SGWB. Since the friction-dominated era of cosmic string dynamics happens right after the formation of the network, and thus in the very early universe, this will result in a contribution in the ultra-high frequency range. As a matter of fact, as illustrated by Fig.~\ref{fig:SGWBfull}, the inclusion of the GWs emitted by the loops that are produced during the friction era extends the range of the spectrum significantly, typically leading to a prominent bump at very high frequencies that may be regarded as a secondary peak. Since friction is often assumed to completely suppress the emission of loops, it is generally assumed that cosmic string networks start contributing to the SGWB in the end of the friction era (at $t=t_f$). As a result, a cut-off is introduced artificially in the high-frequency range of the spectrum and this secondary peak is neglected. However, as the results of Sec.~\ref{sec:fricloop} show, the GW emission of loops during the friction era, although partially reduced, is not completely suppressed (especially during the end stages of the life of the loops and/or towards the end of this era). This partial reduction of the GW emission in many instances does not lead to a complete suppression of the SGWB, since a large number of loops may be created during the friction era. This is well illustrated by Fig.~\ref{fig:nofriction}, where we plot the SGWB sourced during the friction era with and without the inclusion of the suppression of GW emission caused by friction (the latter case was also briefly discussed in~\cite{Gouttenoire:2019kij}). Therein one may see that, although this suppression may indeed lead to a significant reduction of the amplitude of the friction signature on the SGWB, the increase in the number of loops may be so significant that this signature may still survive.

\begin{figure}
	\begin{center}
		\centering
		\includegraphics[width=\linewidth]{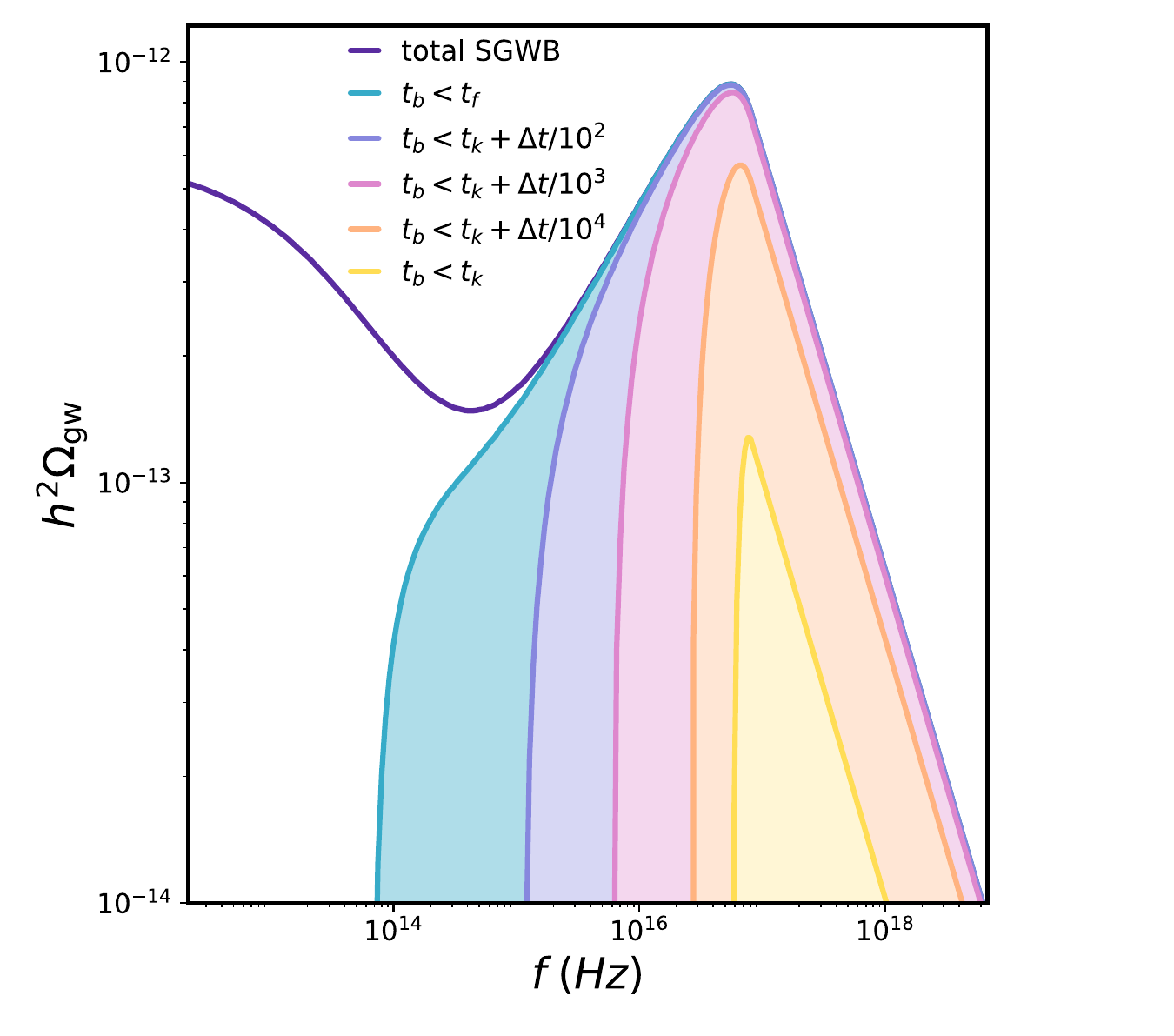}
		\caption{The signature of friction on the SGWB generated by a cosmic string network with $G \mu = 10^{-10}$, $\alpha = 10^{-9}$, $\beta = 1$, $L_c=t_c$. The shaded areas with different colors correspond to the contributions to the spectrum generated by loops that where produced before different instants of time in the friction-dominated era. Here, $\Delta t= t_f-t_k$.}
		\label{fig:SGWBsplitted}
	\end{center}
\end{figure}

The signature of friction is, as shown in Fig.~\ref{fig:SGWBsplitted}, generated mostly in the Kibble regime and, in fact, the contribution of the loops created in the Stretching regime may be considered generally negligible. During the latter, cosmic strings are so effectively damped that their velocities are extremely small, which makes collisions and intercommutations unlikely. In the beginning of the Kibble regime, however, not only the RMS velocity is somewhat larger but the networks are also denser (in comoving coordinates), resulting in loops being profusely produced. Because of that, the dominant contribution to the peak generated by friction is, in fact, sourced in the earlier stages of the Kibble regime --- when the number of loops produced per unit volume is quite large and friction is still quite strong --- and not when friction is at its weakest. This shows that the number of loops produced may, in some instances, be the determinant aspect in having a signature of friction and not how strong the suppression of GW emission is.

\subsection{Impact of cosmic string parameters on the signature of friction}

As we have seen, a clear signature of friction is expected if the loops created in the early stages of the Kibble regime, when the network is very dense and copious amounts of loops are created, lose a considerable amount of energy in the form of gravitational radiation. This signature should then depend not only on the size of loops, parameterized by $\alpha$, but also on the large-scale properties of the cosmic string network during this regime, which are determined by cosmic string tension, the initial characteristic length, and the strength of friction (characterized by $\beta$). In this section, we investigate these dependencies in detail. Since the friction-dominated era happens early in the evolution of the cosmic string network (and, in fact, quite early in cosmic history) and the characteristic lengthscale may be quite small at this stage, some care must be taken to avoid considering unphysical scenarios. Here, we only consider the potential emissions of the loops that are created with lengths that are larger than the Planck length, $l_{pl}$, and assume that GW emission starts only once the gravitational backreaction scale is well defined, with $\Gamma G\mu L > l_{pl}$.

\begin{figure} [h]
     \centering
     \includegraphics[width=\linewidth]{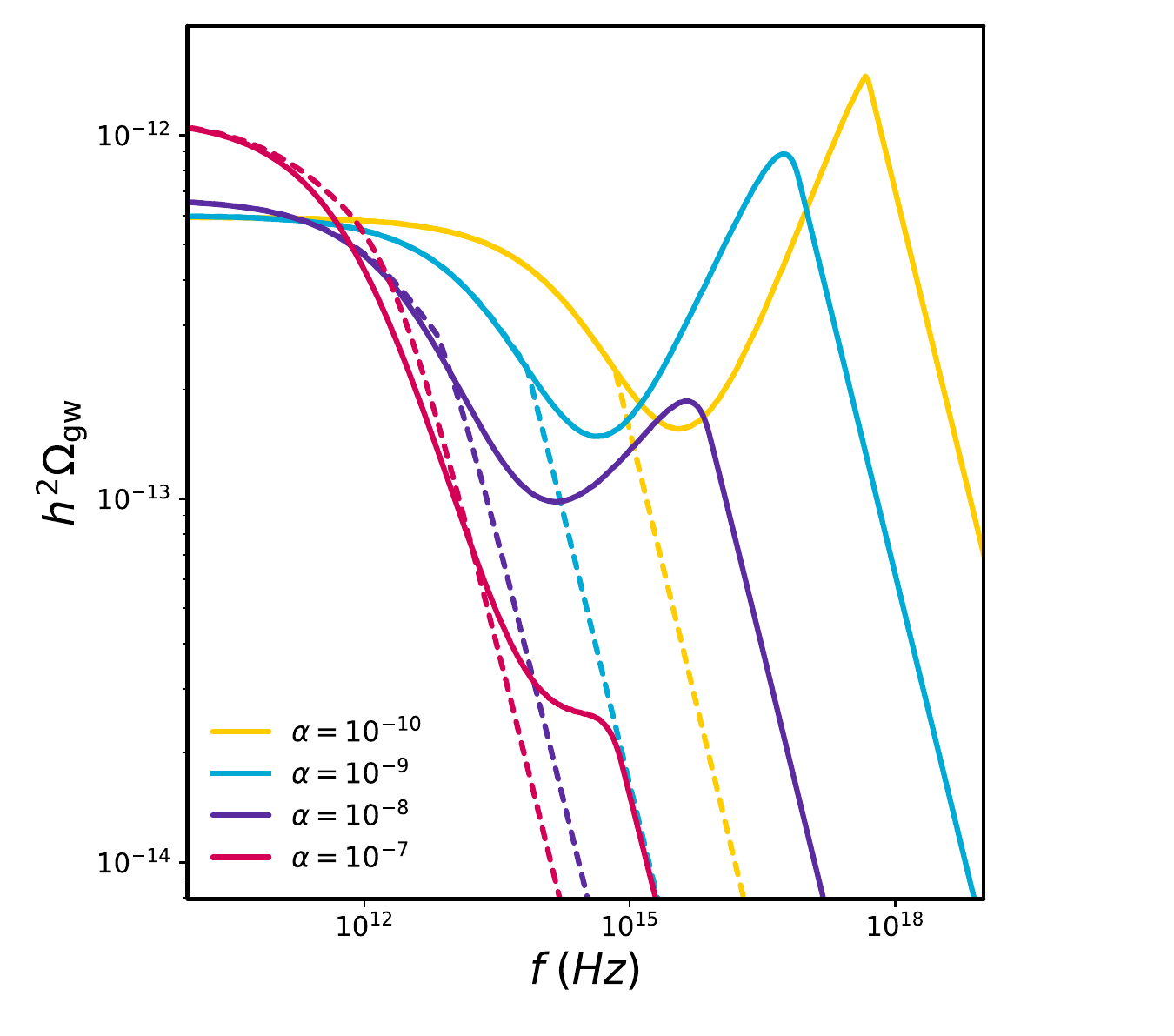}
     \caption{The SGWB generated by cosmic string networks with $G \mu=10^{-10}$, $\beta=1$, $L_c=t_c$ for different values of $\alpha$. The solid lines represent spectra with friction, while the dashed lines correspond to spectra without friction.}
     \label{Fig:alpha}
\end{figure}
\begin{figure} [h]
     \centering
     \includegraphics[width=0.98\linewidth]{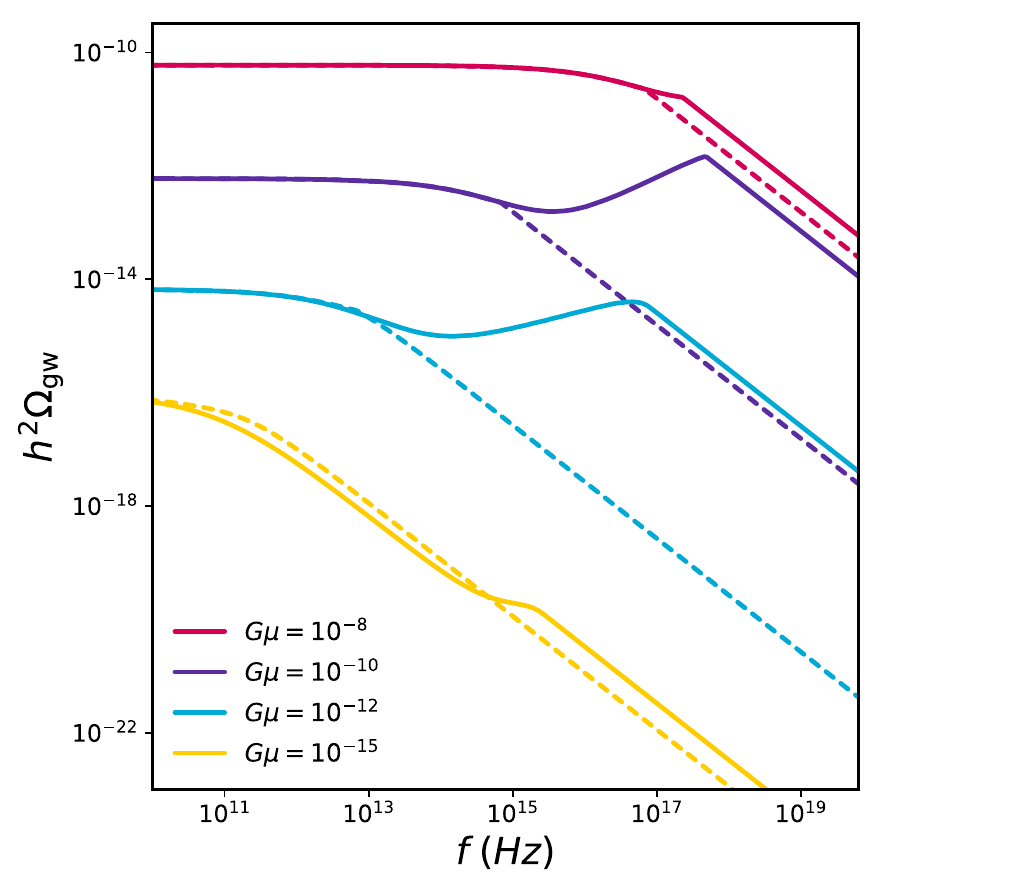}
     \caption{The SGWB generated by cosmic string networks with $\alpha=10^{-10}$, $\beta=1$, $L_c=t_c$ for different values of $G\mu$. The solid lines represent spectra with friction, while the dashed lines correspond to spectra without friction.}
     \label{Fig:Gmu}
\end{figure}
\begin{figure} [h]
     \centering
     \includegraphics[width=\linewidth]{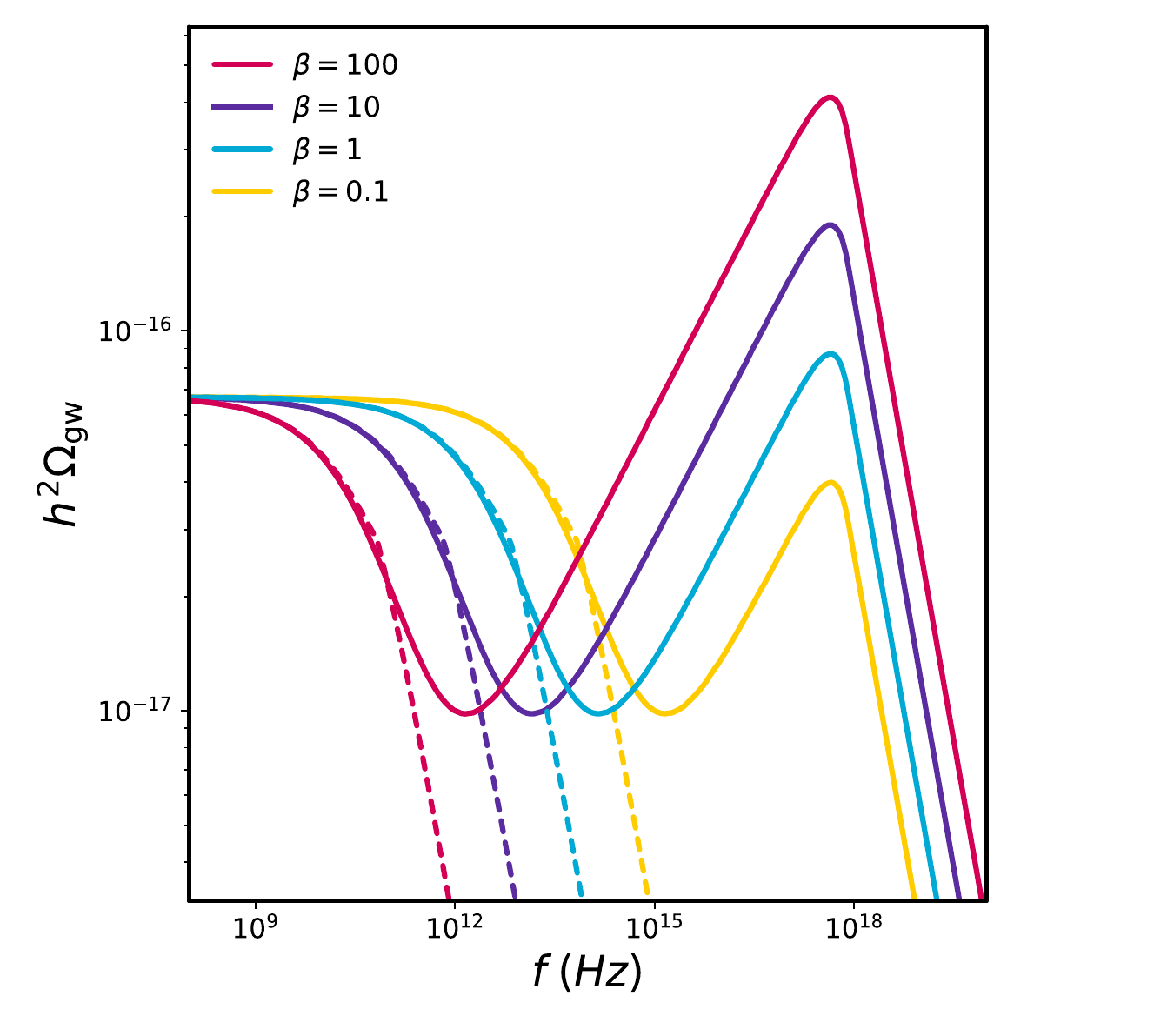}
     \caption{The SGWB generated by cosmic string networks with $G \mu=10^{-14}$, $\alpha=10^{-12}$, $L_c=t_c$ for different values of $\beta$. The solid lines represent spectra with friction, while the dashed lines correspond to spectra without friction.}
     \label{Fig:beta}
\end{figure}
\begin{figure} 
     \centering
     \includegraphics[width=\linewidth]{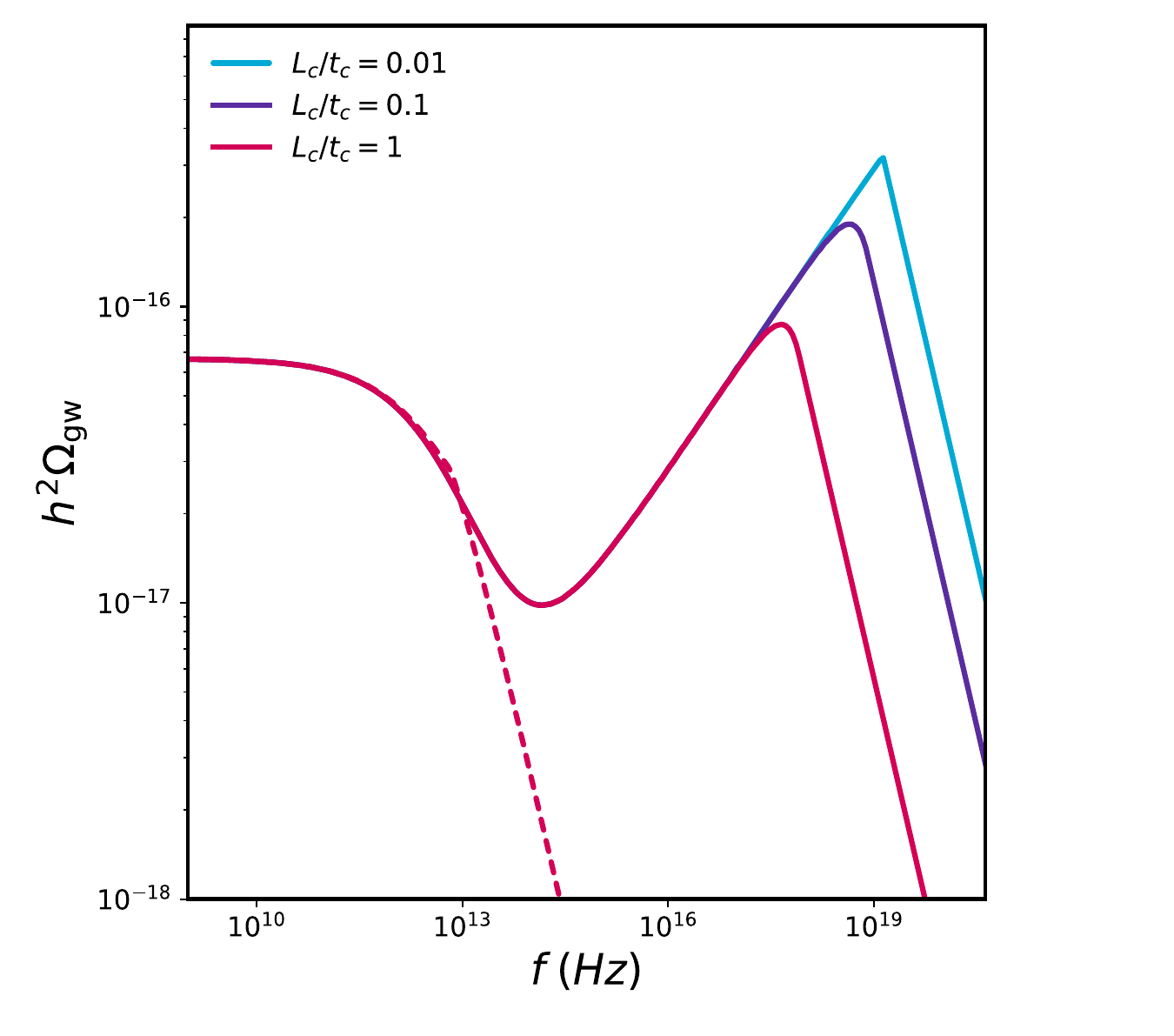}
     \caption{The SGWB generated by cosmic string networks with $G \mu=10^{-14}$, $\alpha=10^{-12}$, $\beta=1$ for different values of $L_c$. The solid lines represent spectra with friction, while the dashed line correspond to spectra without friction.}
     \label{Fig:Lc}
\end{figure}

Let us start by considering the impact of the loop-size parameter $\alpha$ on the signature of friction. As previously explained, as loops evolve and their length decreases, they eventually reach a regime in which their dynamics becomes essentially frictionless. This happens, roughly, once their length becomes smaller than the friction lengthscale and, therefore, loops with a smaller $\alpha$ will reach this regime earlier and consequently lose a larger portion of their energy in the form of GWs. This generally results in a stronger signature of friction as $\alpha$ decreases, as is clearly shown in Fig.~\ref{Fig:alpha}, and this signature is at its strongest in the small-loop regime, with $\alpha\ll \Gamma G\mu$. In this case, in fact, GW emission leads to the evaporation of the loops effectively immediately on cosmological time scales~\cite{Sousa:2014gka} and, as a result, they are barely affected by friction~\footnote{Note that, in the small-loop regime, if $\alpha$ is decreased further, the shape and amplitude of the spectrum is no longer affected and the spectrum is merely shifted towards higher frequencies (see e.g.~\cite{Sousa:2013aaa,Sousa:2014gka}). Recall, however, that loops cannot be made to be arbitrarily small.}. The secondary peak caused by friction is then only present for small enough loops, but how small these loops actually need to be also depends, as we shall see, on the other parameters of the model. In this particular case, the peak starts to form only when loop size approaches the gravitational backreaction scale. However, for larger loops, there is also a signature of friction, although more subtle: the transition between the cut-off of the spectrum and the linear-scaling radiation era plateau is somewhat slower than in the frictionless case. Note also that at the beginning of this transition the spectrum without friction exceeds the spectrum with friction. This is explained by the fact that, around $t_f$, loops still experience a weak frictional force that is not taken into consideration in the computation of the spectrum without friction, which causes a slight suppression of their GW emission (recall that $t_f$ corresponds to the instant of time in which the Hubble and friction damping terms become comparable).

The impact of varying cosmic string tension is slightly more complex. As tension is decreased and strings become lighter, friction damps strings more efficiently (see e.g. Eq.~(\ref{2.3})). Although one would naively expect this to lead to a progressive suppression of this signature, the network also becomes significantly denser during the Kibble regime --- as may be seen in Eq.~(\ref{L_Kibble}) --- and the number of loops then increases with decreasing $G\mu$, leading to a more prominent signature instead. Moreover, GW emission becomes slower when we decrease $G\mu$, which means that, for a fixed $\alpha$, we are moving further into the large-loop regime, in which the signature becomes less visible. These contradictory effects may be seen in Fig.~\ref{Fig:Gmu}, which shows the signature of friction in the SGWB for different values of $G\mu$ and a fixed $\alpha$. Although, initially a decrease of tension leads to an increase of the relative height of the friction peak~\footnote{The very pointy peak and sharp cut-off seen in the spectra with $G\mu=10^{-10}$ and $G\mu=10^{-8}$ are caused by the conditions imposed to ensure that gravitational backreaction scale and the length of loops are physical and well defined. We have, however, verified that the behaviour is qualitatively similar if these conditions are not enforced.}, the friction signature starts to become less prominent as the tension is decreased further. Moreover, since $t_k\sim (G\mu)^{-4/3}$, decreasing of the tension also causes a shift of the friction peak towards lower frequencies and the Kibble regime actually lasts longer (since $t_f\sim(G\mu)^{-2}$), thus affecting a broader range of frequencies.

Increasing $\beta$, which depends on the number of particle species that interact with the strings and thus quantifies the strength of friction, increases the duration of the friction era. As a result, the friction peak is broadened and the SGWB spectrum is affected by friction up to lower frequencies, as Fig.~\ref{Fig:beta} shows. However, an increase in $\beta$ also means that the network becomes denser during the Kibble regime, leading to an increase of the height of the friction peak, and that this regime happens at later times, causing the very slight shift of the peak frequency that may be seen in this figure.

Finally, as discussed in Sec.~\ref{sec:friction-network}, if one decreases the initial characteristic length of the network, the Kibble regime, in which this signature is created, will start earlier and the network will be much denser in this stage. This leads, as shown in Fig.~\ref{Fig:Lc}, to a significant increase of the height of the peak and to an extension of its range towards even higher frequencies. Note however that for very small $L_c$, $\Gamma G\mu L$ and $\alpha L$ necessarily become unphysically small at the very early stages of the evolution of the network.

\section{Discussion}\label{sec:discussion}

We have shown that friction may leave a distinct signature, in the form of a secondary peak, in the ultra-high frequency range of the SGWB spectrum generated by cosmic string loops. This signature has been neglected in the literature, since it is usually assumed that friction leads to a very strong suppression of the GW emission of cosmic string loops. We found, however, that the end stages of the life of loops will necessarily be frictionless and that their emission is only partially reduced. As a matter of fact, during the Kibble regime, when the network is very dense and intercommutations are very frequent, loop production is very intensive and the superposition of their numerous individual emissions despite being weakened may still generate a prominent characteristic imprint.

The signature of friction uncovered here may potentially allow us not only to probe cosmic string tension (and through it the energy scale of the string-forming phase transition), but also the details of the underlying particle physics scenario. As we have seen, the height and broadness of the secondary peak caused by friction depends on the number of particle species of the surrounding plasma that interact strongly with the strings, quantified through the parameter $\beta$. This parameter necessarily depends on the nature of the fields that constitute the string and, therefore, one may expect different string models to generate distinguishable signatures. This signature is also sensitive to the initial conditions of the cosmic string network, which may also allow us to uncover more about the symmetry breaking process that originated the strings. As a matter of fact, an initial characteristic length of the order of the cosmological horizon would be expected in slow first order phase transitions, while smaller $L_c$ (up to $L_c \sim t_c /\theta$~\cite{Martins:1996jp}) may result from second-order phase transitions~\cite{Vilenkin:2000jqa,Martins:1996jp}. Note that the evolution of the cosmic string network in the frictionless regime is insensitive to these two aspects and, therefore, this signature of friction offers the possibility of extracting information about the physics of the early universe that the rest of the SGWB does not allow us to access.

Our results show that the existence of this signature is also strongly dependent on the size of loops and, in fact, this signature is more prominent roughly when the length of loops is not much larger than the gravitational backreaction scale (but this, as we have seen, depends on a number of other aspects too). Nambu-Goto numerical simulations~\cite{Blanco-Pillado:2013qja} indicate that 10\% of the energy lost by the network as a result of interactions is in the form of large loops with $\alpha \sim 0.34$ and that the rest of the energy is lost in the form of small loops with $\alpha\sim \Gamma G\mu$ and thus the majority of the loops created is, in fact, quite small. So, although the contribution of the large loops (that may provide the dominant contribution to the SGWB in the frictionless regime) would be highly suppressed in the friction era, the numerous small loops could still give rise to a prominent signature in the ultra-high frequency range. Note that loop production in the presence of friction has not yet been studied in detail and there is the possibility that the production of small loops may be suppressed in these early stages of the evolution, since friction could potentially erase small-scale structure on long strings on time scales smaller than a Hubble time~\cite{M_B_Hindmarsh_1995,Vilenkin:1981bx}. However, since during the Kibble regime collisions and intercommutations between strings are quite frequent, new kinks should be copiously produced during this regime. Friction should also only be effective in removing structure on scales that are larger than the friction lengthscale. Whether the kink decay caused by friction is effective in smoothing long strings and in suppressing the production of small loops is then a subject that warrants further investigation.

We also noticed that unphysically small scales for the length of loops and for the gravitational backreaction scale may arise during the friction epoch. Since this could mean that the classical treatment of cosmic strings might not be well justified anymore and that quantum effects may have to be considered (if thermal fluctuations are not large enough to prevent the existence of strings at these scales), we have introduced cut-offs to address these situations in our computations. Quantum fluctuations on straight strings in Minkowski spacetime in the weak coupling limit were studied in~\cite{Garriga:1991tb} and the results indicate that these fluctuations are generally small enough, when compared to string thickness, to be neglected. However, if quantum corrections are indeed necessary, in principle, they may leave a signature in the SGWB in the ultra-high frequency range (which would appear in the spectra where the cut-off we have imposed have a visible effect).

A detection of the signature of friction in the ultra-high frequency range of the SGWB spectrum is currently challenging. This range of the spectrum has, however, been garnering a lot of attention in the literature and several observational concepts to probe it have been proposed (see e.g.~\cite{Aggarwal:2020olq} for a recent review). Given the location of the signature, experiments based on the inverse Gertsenshtein effect~\cite{Gertsenshtein1962, PhysRevD.37.1237}, also known as magnetic conversion, are of particular interest, as they may, in principle, be designed to detect GWs with frequencies above $1$ THz. This effect --- the conversion of GWs into photons in the presence of an external magnetic field --- may also take place in current and upcoming axion-search experiments, in which axion-like particles may convert into photons in the presence of a strong magnetic field. In~\cite{2019EPJC...79.1032E}, the authors use ALPS, OSQAR and CAST data to constrain the amplitude of isotropic ultra-high frequency GWs in two frequency bands: $(2.7-14) \times 10^{14}$ Hz and $(5-12) \times 10^{18}$ Hz. Besides experiments based on artificial magnetic fields, there are suggestions to also use astronomical magnetic fields such as those of pulsars~\cite{Ito:2023fcr}, planetary magnetospheres~\cite{Liu:2023mll} or galaxy clusters~\cite{He:2023xoh} to detect GWs. The former~\cite{Ito:2023fcr} constrain SGWBs around $10^{13}-10^{27}$ Hz, precisely in the relevant frequency range to probe friction. Moreover, in~\cite{Vacalis:2023gdz}, the authors proposed a new method based on the inverse Gertsenshtein effect that resorts to high-energy laser beams instead of magnetic fields. They find that the best sensitivity is reached when the GW frequency is twice the frequency of the lasers used and thus, given the wide range of lasers' operational frequencies ($10^{13}-10^{19}$ Hz), this method may allow us to probe GWs in a broad frequency range in the future. So, although this signature is currently out of reach of GW detectors, the sensitivity to SGWBs in the ultra-high frequency range of the spectrum is expected to increase, thus opening the prospect of probing the imprints of friction directly in the future.

%%%%%%%%%%%%%%%%%%%%%%%%%%%%%%%%%%%%%%%%%%%%%%%%%%%%%
\begin{acknowledgments}
The authors thank Pedro Avelino and Andreas Neitzel for many enlightening discussions. S. M. is supported by FCT  - Funda\c{c}\~{a}o para a Ci\^{e}ncia e a Tecnologia through the PhD fellowship UI/BD/152220/2021. L. S. is supported by FCT through contract No. DL 57/2016/CP1364/CT0001. Funding for this work has also been provided by FCT through the research grants UIDB/04434/2020 and UIDP/04434/2020 and through the R \& D project 2022.03495.PTDC -- \textit{Uncovering the nature of cosmic strings}.

\end{acknowledgments}
%%%%%%%%%%%%%%%%%%%%%%%%%%%%%%%%%%%%%%%%%%%%%%%%%%%%%%%%%%
 
\bibliography{literature}
 	
 \end{document}